\documentclass[showpacs,showemail,amsfonts]{revtex4}  
\usepackage{amssymb}
\usepackage{graphicx}
\usepackage{bm}

\begin{document}

\title[Effects of Low-degree Preferential Attachment] { 
Effects of Preference for Attachment to Low-degree 
Nodes on the Degree Distributions of a Growing Directed Network and a Simple 
Food-Web Model}

\author{Volkan Sevim$^1$}
\email{sevim@scs.fsu.edu}
\author{Per Arne Rikvold$^{1,2}$}
\email{rikvold@scs.fsu.edu}
\affiliation{
$^1$ School of Computational Science,
Center for Materials Research and Technology,
and Department of Physics, 
Florida State University, Tallahassee, FL 32306-4120, USA\\
$^2$ National High Magnetic Field Laboratory,
Tallahassee, FL 32310-3706, USA}
\date{\today}
\newcommand{\nk}{$n_k^*$}
\newcommand{\newzeta}{m/z_m^*}
\newcommand{\zstar}{z_m^*}
\newcommand{\meank}{\left\langle k\right\rangle}

\begin{abstract}
We study the growth of a directed network, in which the growth is constrained 
by the cost of adding links to the existing nodes. We propose a new 
preferential-attachment scheme, in which a new node attaches to an existing 
node $i$ with probability $\Pi(k_{i})\propto k_{i}^{-1}$, where $k_i$ is the 
number of outgoing links at $i$.  We calculate the degree distribution for the 
outgoing links in the asymptotic regime ($t\rightarrow\infty$), \nk, both 
analytically and by Monte Carlo simulations. The distribution decays like 
$k\mu^k/\Gamma(k)$ for large $k$, where $\mu$ is a constant.  We investigate 
the effect of this preferential-attachment scheme, by comparing 
the results to an equivalent growth model with a degree-independent 
probability of attachment, which gives an exponential outdegree distribution. 
Also, we relate this mechanism to simple food-web models by implementing it 
in the cascade model. We show that the low-degree 
preferential-attachment mechanism breaks the symmetry between in- and 
outdegree distributions in the cascade model. It also causes a faster decay 
in the tails of the outdegree distributions for both our network growth model 
and the cascade model.
\end{abstract}


\maketitle

\section{Introduction}
\label{sec:I}
Many real-world networks that have been studied recently, such as the 
internet, 
the World Wide Web, and social relations have a scale-free structure, whose 
degree distribution, $P(l)$, follows a power law for large $l$, the number of 
links connecting to a single node. A scale-free network can be constructed 
using a preferential-attachment mechanism in which a new node attaches to an 
existing node, $i$, which has $l_i$ links with a probability 
$\Pi(l_i)=l_i/\sum_{j}{l_j}$~\cite{Barabasi:1999}. This form of 
preferential attachment means that the highly 
connected nodes attract more new nodes than the others, and it explains the 
structure of networks like the internet, 
in which a connection is favorable for both connected nodes, and where links 
are undirected. Growth according to this mechanism leads to self-organization 
into a scale-free structure~\cite{Barabasi:1999, Barabasi:2002, 
Dorogovtsev:2000, Krapivsky:2000, Krapivsky:2001}.

In this paper, we propose a new preferential-attachment scheme for growing 
directed networks~\cite{Dorogovtsev:2001, Rodgers:2001}, in which new nodes 
prefer to attach to existing nodes with {\it lower} degree. Such a mechanism 
could play a role in some transportation 
networks~\cite{Garlaschelli:2003,Banavar:1999} such as food 
webs~\cite{Drossel:2002,  Martinez:2002, Montoya:2003, McKane-Rev} or 
power-grids~\cite{Amaral:2000, Albert:2004}, which can be considered directed. 
The links in such networks transport some sort of resource to ``feed'' the 
nodes. Each node gets resources only by feeding on another one. (The initial 
nodes can be considered as sources.) We take the direction of the resource 
flow as the direction of the link. Therefore, an outgoing link transports 
resources from a ``prey'' node to a ``predator'' node. If there is more than 
one predator that feeds on a prey node, then the resources supplied by the 
prey have to be shared by the predators. This implies that for a conventional 
growth model, the attractiveness of a node in such a network should increase 
with the number of its incoming links (indegree), and decrease with the number 
of its outgoing links (outdegree). In other words, the cost of adding links to 
an existing node should increase due to limited capacity, as the number of 
nodes that feed on it increases. This cost puts a constraint on the growth of 
the network~\cite{Amaral:2000}. A good example of such a system, the 
world-wide airport network, is studied in Ref. ~\cite{Guimera:2004}.

Therefore, we propose a new growth scheme, in which a new node attaches to an 
existing node, $i$, with probability $\Pi(k'_{i}, k_{i})\propto 
(k'_{i}/k_{i})^{\gamma}$, where $k_{i}$ and $k'_{i}$ are the number of 
outgoing and incoming links at node $i$, respectively. This means that new 
nodes prefer connecting to existing nodes with more incoming and fewer 
outgoing links because they provide more resources per outgoing link. Here we 
study a special case in which $k'_i = m$ for all nodes.

As mentioned above, we believe that such a mechanism could play a role in 
real systems such as food webs, which provide a good basis for testing our 
predictions. There has been significant progress in this field in the last 
decade, especially in development and understanding of static models, which 
have been quite successful at reproducing webs with structural properties in 
agreement with the empirical data~\cite{Drossel:2002, McKane-Rev, 
Martinez:2000, Martinez:2002, Camacho:2002A, Camacho:2002B, Stouffer:2005, 
Montoya:2003}. Despite their success, most, if not all, proposed static 
food-web models in the literature such as the cascade 
model~\cite{CommunityFoodWebs} and the niche model~\cite{Martinez:2000} are 
phenomenological models. They do not employ physical principles in the 
assembly process --- or at least, the rules they employ do not yet have any 
physical 
interpretations. Although the mechanism we propose in this paper is quite 
na{\'\i}ve, we believe that a static food-web model should employ such rules, 
which can be related to physical principles.

The structure of the paper is as follows. We describe the model, derive the 
outdegree distribution, and compare the results with Monte 
Carlo simulations in Sec.~\ref{sec:M}. Some details of the calculations are 
given in Appendix~\ref{sec:zstar}. In Sec~\ref{sec:role}, 
we compare the proposed model to an equivalent model with a 
degree-independent probability of attachment. The derivation of the outdegree 
distribution for this model is provided in Appendix~\ref{sec:const}. In 
Appendices~\ref{sec:app} and~\ref{sec:const}, we show that the outdegree 
distributions for these growth models can be scaled onto a single scaling 
function for large $m$. In Sec~\ref{sec:role}, we also relate our results to 
some food-web models. We summarize our results in Sec.~\ref{sec:sum}. 
\begin{figure}[t]
\centering \vspace{0.6truecm} 
\includegraphics[width=0.35\textwidth]{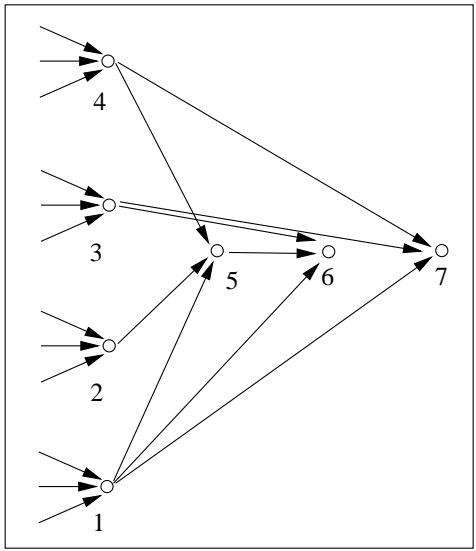}
\caption{ Illustration of the growth with $N_0=4$ and $m=3$. Nodes 1, 2, 3, 
and 4 are the initial (source) nodes connected to unspecified nodes outside 
the network. Nodes 5, 6, and 7 are the nodes added at timesteps 1, 2, and 3, 
respectively. See text for details. } 
\label{fig:illustration}
\end{figure}

\section{Model and Results}
\label{sec:M}
The growth begins with $N_0\geq m$ nodes with $m$ incoming links, each 
connected to unspecified nodes outside the network. These initial nodes can be 
considered as sources, and the only purpose of their links is to make sure 
that each node has the same number of incoming links. We add one node with $m$ 
incoming links at each time step, which connects to $m$ different nodes. The 
direction of the new links are the same as the direction of the resource flow: 
from the existing node to the new one. An example of such a network after the 
first few time steps is shown in Fig.~\ref{fig:illustration}.

As we stated above, an existing node acquires a new link with probability 
$\Pi(k'_{i}, k_{i})\propto (k'_{i}/k_{i})^{\gamma}$. However, for the case in 
which all nodes have $m$ incoming links, $k'_{i}=m$ is just a scale factor. 
Without changing the proportionality for $k_{i}\gg 1$, we replace $k_i$ by 
$k_i+1$ in order to keep $\Pi(k_{i})$ finite for $k_i = 0$. Here, we study a 
special case of this attachment probability, $\Pi(k_{i})\propto k_{i}^{-
\gamma}$, in which $\gamma=1$. Therefore, the probability of attachment 
is given by
\begin{equation}
\Pi(k_{i}) = \frac{1}{(k_{i}+1)Z_m}
\;,
\label{eq:attachprob}
\end{equation} 
where 
\begin{equation}
Z_m=\sum_{i=0}^{N_{\rm T}}{\frac{1}{k_{i}+1}} = 
\sum_{k=0}^{\infty}{\frac{N_{k}}{k+1}}
\;.
\end{equation} 
Here, $N_{k}$ is the number of nodes with outdegree $k$, and $N_{\rm T}$ is the 
total number of nodes. We note that $N_{k}$, and therefore, $Z_m$, depends on 
time.

We use the rate-equation approximation of 
Refs.~\cite{Krapivsky:2000, Krapivsky:2001}. In the limit of $N_{\rm T}\gg 1$, 
the 
rate equations for the outgoing links are:
\begin{equation}
\frac{dN_0}{dt} = 1-m\Pi(0)N_0\;,
\end{equation}
\begin{equation} 
\frac{dN_{k}}{dt} = m\Pi(k-1)N_{k-1} - m\Pi(k)N_{k}
\qquad\textrm{for }k>0
\;.
\end{equation} 
Substituting  $n_{k}N_{\rm T}$ for $N_{k}$, and using $dN_{\rm T}/dt=1$, yields 
corresponding equations for the density of nodes:
\begin{equation}
\label{eq:n0t}
\frac{dn_0}{dt} = \frac{1-n_0}{N_{\rm T}} - m\Pi(0)n_0 
\;,
\end{equation} 
\begin{equation}
\label{eq:nkt}
\frac{dn_{k}}{dt} 
	= m\Pi(k-1)n_{k-1} - n_{k}\left[{N_{\rm T}^{-1}}+m\Pi(k)\right]
\qquad\textrm{for }k>0
\;.
\end{equation}
We are interested only in the asymptotic regime ($t\rightarrow\infty$), in 
which $dn_{k}/dt=0$, and the distribution reaches a steady state. 
The steady-state solutions of Eqs.~(\ref{eq:n0t}) and~(\ref{eq:nkt}) are 
\begin{equation}
\label{eq:n0}
n_0^*=\frac{1}{1+\newzeta} \;,
\end{equation}
\begin{equation}
\label{eq:nk}
n_{k}^*=\frac{n_{k-1}^*}{k[(k+1)^{-1}+ z_m^*/m]} 
\qquad\textrm{for }k>0
\;,
\end{equation} 
where 
\begin{equation}
\label{eq:z}
z_m^*=\lim_{t\to\infty}Z_m/N_{\rm T} = \sum_{i=0}^{\infty}{\frac{n_{i}^*}{i+1}} 
\;.
\end{equation}
The solution of the recursion relation Eq.~(\ref{eq:nk}) can be obtained by a 
few straightforward algebraic manipulations. First, we convert the right-hand 
side of Eq.~(\ref{eq:nk}) into a product, using Eq.~(\ref{eq:n0}):
\begin{equation}
n_k^*=\frac{(\newzeta)^k (k+1)}{1+\newzeta}\prod 
_{j=1}^{k}\frac{1}{j+1+\newzeta}
\qquad\textrm{for }k>0
\label{eq:nkprod}
\;.
\end{equation}
Then, using the identity $\Gamma(x+1)=x\Gamma(x)$, we write the product in 
terms of a ratio of Gamma functions to obtain the final form of the outdegree 
distribution:
\begin{equation}
\label{eq:fullnk}
n_{k}^*=(k+1)(\newzeta)^{k}\frac{\Gamma \big(1+\newzeta\big)}{\Gamma 
\big(k+2+\newzeta\big)}
\qquad\textrm{for }k\geq0
\;.
\end{equation}
We note that Eq.~(\ref{eq:fullnk}) holds for $k=0$ as well. (The outdegree 
distribution, Eq.~(\ref{eq:fullnk}), is analogous to the predator distribution 
in a food web. We use the terms in- and outdegree distribution interchangeably 
with prey and predator distribution, respectively.) The outdegree distribution, 
\nk, is shown in Fig.~\ref{fig:nkplot} for $m=1,3,$ and 5. The method we use
to obtain $\zstar$ is explained in Appendix~\ref{sec:zstar}. Also, in Appendix
~\ref{sec:app}, we prove that the scaled distributions 
$mn_k^*$ vs $u=k/m$ collapse onto a single scaling function, 
\begin{equation}
\label{eq:scaling}
p(u)=\frac{\pi u}{2}\exp (-\pi u^2/4)\;, 
\end{equation}
for large values of $m$.

\begin{figure}[t]
\centering \vspace{0.6truecm} 
\includegraphics[width=8.5truecm]{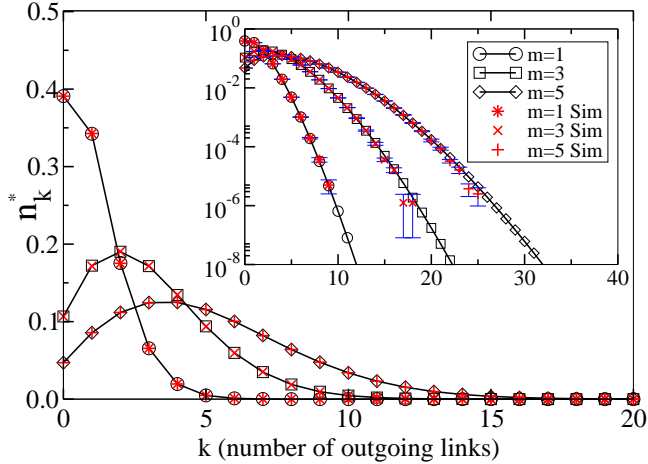}

\caption{(Color online) Outdegree (predator) distributions, $n_k^*$, for the 
growth model 
with low-degree preferential attachment, for $m=1$, 3, and 
5 with $N_0=10$. The results of the Monte Carlo simulations (the symbols 
$\ast$, $\times$, and $+$) are in excellent agreement with the theory (the 
symbols $\circ$, $\Box$, and $\diamond$). As $k$ is a discrete variable, the 
lines connecting the symbols are merely guides to the eye. The simulations 
were performed up to a system size of $N_{\rm T}=N_0+10^6$ nodes, and averaged 
over 
eight independent runs. Inset: The same distributions shown on a log-linear 
scale. The tails decay faster than an exponential, indicating that the 
topology of the network is relatively homogeneous.} 
\label{fig:nkplot}
\end{figure}

\begin{figure}[t]
\centering \vspace{0.6truecm} 
\includegraphics[width=8.5truecm]{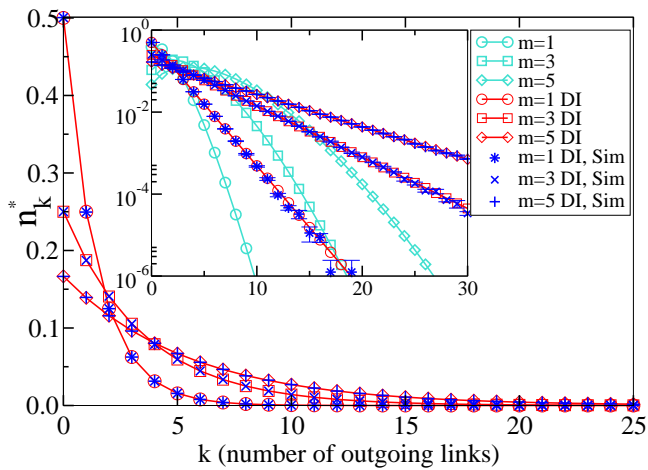}
\caption{(Color online) 
Outdegree (predator) distributions, $n_k^*$, for the equivalent growth 
model with degree-independent probability of attachment, for $m=1$, 3, and 
5 with $N_0=10$. The results of the Monte Carlo simulations (the symbols 
$\ast$, $\times$, and $+$) are in excellent agreement with the theory, which 
predicts exponential decay (the symbols $\circ$, $\Box$, and $\diamond$). 
Compare to the degree-dependent case in Fig.~\ref{fig:nkplot}. The simulations 
were performed up to a system size of $N_{\rm T}=N_0+10^6$ nodes, and averaged 
over 
eight independent runs. Inset: The same distributions shown on a log-linear 
scale along with the theoretical results for the degree-dependent case (gray 
curves - turquoise online) in Fig.~\ref{fig:nkplot} for comparison. } 
\label{fig:nk-degree-indep}
\end{figure}

\begin{figure}[t]
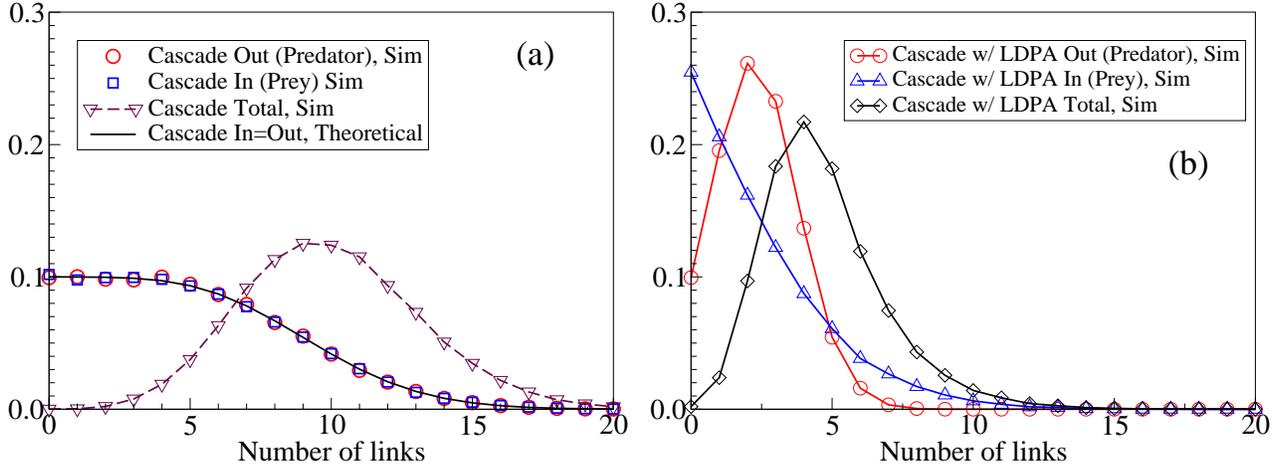

\centering
\includegraphics[width=.47\textwidth,viewport= 0mm 0mm 241mm 175mm ,clip]
{cascade-new.eps}
\includegraphics[width=.47\textwidth,viewport= 0mm 0mm 241mm 175mm ,clip]
{cascade-w-PA-new.eps}	\\
\vspace{0.1truecm}
\caption[]{(Color online) Degree distributions 
for the cascade model with and without low-degree preferential attachment with 
$d=10$ and $S=10^4$. The error bars for the Monte Carlo data (averaged over 
eight runs) are  smaller than the symbols. ``Total'' represents the 
distribution for the total number of links, $k+k'$. (a) The distributions for 
the 
original cascade model, in which the probability of attachment is $x=d/S$ (the 
symbols, $\circ$, $\Box$ and $\bigtriangledown$). The in- and outdegree 
distributions obtained from the Monte Carlo simulations practically overlap in 
agreement with the theory, Eq.~(\ref{eq:cascadedegree}) (black line).  (b) The 
distributions for the cascade
model with low-degree preferential attachment with $x(k_i)=d/[S(k_i+1)]$. The 
in-outdegree symmetry of the original cascade model is broken in the new 
distributions. See text for details.} 
\label{fig:cascade}
\end{figure}

\begin{figure}[t]
\includegraphics[width=.47\textwidth,viewport= 0mm 0mm 250mm 180mm ,clip]
{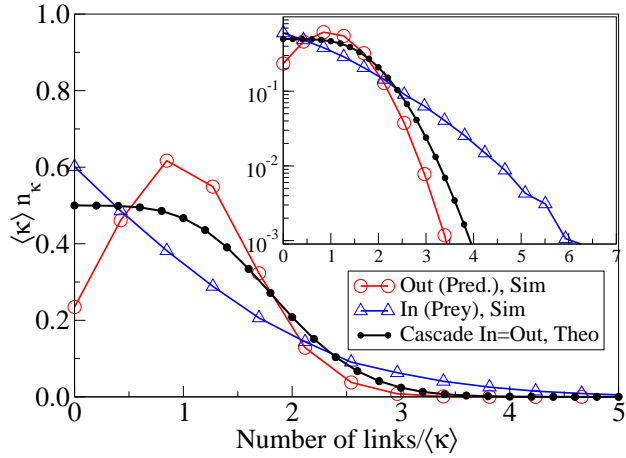}		
\caption[]{(Color online) Scaled in- and outdegree distributions for the 
cascade model with (the symbols, $\bigtriangleup$ and $\circ$) and without 
(solid circles) low-degree preferential attachment. In the axis titles, 
$n_\kappa$ represents the degree distributions, \nk, $n_{k'}^*$, and 
$p^{(in=out)}(k)$, and $\left\langle \kappa \right\rangle$ represents their 
respective mean values, which are 5 for the original cascade model, and 2.36 
for the modified model. Each distribution is scaled using its respective 
$\left\langle \kappa \right\rangle$. The in- and outdegree distributions have 
the same mean for each case. Inset: The same distributions shown on a 
log-linear scale for comparison.} 
\label{fig:cascade-rescaled}
\end{figure}

We also check our theoretical results by comparing them to results of Monte 
Carlo simulations. The growth scheme in the simulations is implemented as 
follows. We begin the simulation with $N_0\geq m$ nodes with $m$ incoming 
links each. At each time step, we add one new node. We pick a random existing  
node from 
the network, calculate the probability of attachment to that node, and 
generate a random number. If the random number is smaller than the attachment 
probability, we create a link between the new node and the randomly picked 
node. If not, we randomly pick another node from the network and repeat the 
process until the new node has $m$ links. (We do not pick the same node 
twice.) The direction of a new link is from the existing node to the new one. 
The simulation stops when the system size reaches $N_{\rm{max}}+N_0$ nodes. We 
take $N_{\rm{max}}=10^6$ and average the results over eight independent runs.

The results of the Monte Carlo simulations are also shown in 
Fig.~\ref{fig:nkplot}. 
They are in excellent agreement with the theory. The deviations in the last 
three nonzero data points are consistent with the number of runs used for 
averaging.

In order to clarify the role of the low-degree preferential attachment, we 
provide some comparisons with similar models in the next section. 


\section{The Role of the Low-degree Preferential Attachment and Relevance to 
Simple Food-Web Models}
\label{sec:role}
\subsection{A growing network with degree-independent probability of 
attachment}
We first compare our model to an
 equivalent model with a degree-independent probability of attachment, 
$\Pi=1/N_{\rm T}$. 
In this case, $\Pi$ is the same for all nodes, however, it is time dependent. 
As we prove in Appendix~\ref{sec:const}, a degree-independent $\Pi$ yields 
a network with an exponential outdegree distribution as seen in 
Fig.~\ref{fig:nk-degree-indep}. This shows that the low-degree
preferential attachment is responsible for the peaked form of \nk, and also 
for the faster decay in the tail.

\subsection{The cascade model}
Secondly, we test the effect of the low-degree preferential attachment in the 
cascade model~\cite{CommunityFoodWebs, Stouffer:2005}, which is a simple 
wiring-based food-web 
model. Although the cascade model has now mostly been replaced by 
improved variants like the niche model~\cite{Martinez:2000}, it serves our 
purpose better, because it is quite simple 
and transparent. The cascade model builds a web using a few simple rules as 
follows. The species are numbered from 1 to $S$, where $S$ is the 
total number of species. Each 
species, $i$, is allowed to prey on a species $j$, where $i>j$, with a 
constant 
probability $x=d/S$, where $d\le S$ is a constant. The species with no 
predators constitute the top species, and the species with no prey constitute 
the basal species
~\footnote{There is also a small probability of creating isolated species. The 
probability that a species $i$ has no links is 
${\cal P}_i^{\rm (total)}(k=0)=(1-d/S)^{S-1}$. Therefore, the expected number 
of species with no links is $S(1-d/S)^{S-1}$, which can be approximated as 
$S\exp(-d)$ for $S\gg d$.}. The average number of prey (or 
predator) per species is equal to $(d/2)(1-1/S)\approx  d/2$ for large $S$. 
Therefore, $d/2$ is analogous to $m$ in our growth model.

The calculation of the outdegree (predator) distribution of the webs produced 
by the 
cascade model is quite straightforward. First, we find the probability of 
finding a node, $i$, with $k$ outgoing links:
\begin{equation}
{\cal P}_i^{\rm (out)}(k) = {S-i \choose {k}} (1-x)^{S-i-k} x^k
\qquad\textrm{with } k\leq S-i \;.
\end{equation}
Then we obtain the outdegree distribution by summing 
${\cal P}_i^{\rm (out)}(k)$ over $i$, and normalizing by $S$ 
\footnote{The in-outdegree distributions we obtain for the cascade model are 
same as the outdegree distribution for the niche model as derived by Camacho et 
al. in Ref.~\cite{Camacho:2002A}, except that they use a mean value for $x$.}:
\begin{equation}
\label{eq:cascadedegree}
p^{\rm (out)}(k) = \frac{1}{S}\sum_{i=1}^{S-k}{{\cal P}_i^{\rm (out)}(k)}
\qquad\textrm{with } k<S \;.
\end{equation}
The cascade model produces webs with identical in- and outdegree 
distributions. One can prove this by substituting ${\cal 
P}_{S-i+1}^{\rm (out)}(k)$ for ${\cal P}_i^{\rm (in)}(k)$ in 
$p^{\rm (in)}(k) = \sum_{i=k+1}^{S}{{\cal P}_i^{\rm (in)}(k)}/S$.

In order to see the effect of the low-degree preferential attachment, we modify 
the cascade model by changing 
the probability of attachment from a constant, $x=d/S$, to a decreasing 
function of the outdegree of the prey node, $x(k_i)=d/[S(k_i+1)]$. (We note that 
for both original and modified cascade models, the probability of attachment 
does not depend on time, unlike in our proposed growth model.) Obviously, 
this modification does not conserve the expected total number of links. 
Nevertheless, it still demonstrates the effect 
of the low-degree preferential attachment in the cascade model. We also point 
out that the degree-dependent nature of this mechanism could cause 
different degree distributions in the cascade model, depending on the order 
in the wiring procedure. (See endnote ~\footnote{ One could begin 
the wiring from the species with the highest (lowest) species index and 
proceed to the bottom (top), or pick the species at random. For the original 
cascade model this order is irrelevant. However, if preferential 
attachment is involved, this order could affect the distributions, since this 
mechanism is degree-dependent. We use a random order of assembly in 
our simulations to avoid a possible bias: 
a species, $i$, is picked at random, and all species with a 
lower index are linked to $i$ with a probability as described in the text.}.) We 
leave extensive analysis to a further publication and give only these 
preliminary results here. 

Fig.~\ref{fig:cascade} shows the in- and outdegree distributions of the 
original and modified
cascade models for $d=10$ and $S=10^4$. One can clearly discern two changes: 
the new in- and outdegree distributions are not identical, and they are also 
quite different than those of the original cascade model. The new outdegree 
distribution takes a similar form to the outdegree 
distribution in our model, $n_k^*$, in Eq.~(\ref{eq:fullnk}), and the new 
indegree distribution decays monotonically. For a better comparison, we also 
provide the scaled forms of these distributions in 
Fig.~\ref{fig:cascade-rescaled}. The figure shows that the tail of the 
outdegree distribution is depressed, while the indegree distribution decays 
much slower than in the original cascade model. All the distributions decay 
faster than an exponential. The analytical forms of the distributions are 
left for further investigation.

One could also compare our results with the latest findings in food-web 
theory. According to Camacho et al. \cite{Camacho:2002A, Camacho:2002B, 
Stouffer:2005}, the predator and prey distributions for most empirical food 
webs obey universal functional forms. The universal form of the predator 
(outdegree) distribution is given by $p_{\rm pred}(k)= \gamma(k+1,z)/z$, where 
$\gamma(k,z)$ is the incomplete Gamma function 
and $z$ is the average connectivity, $2L/S$, with $L$ the total number 
of links in the web. This is a continuum approximation to 
Eq.~(\ref{eq:cascadedegree}) for $S\gg 1$ and $x\ll 1$. 
This form of the predator 
distribution suggested by Camacho 
et al. and the outdegree distribution of our model with low-degree preferential 
attachment, $n_k^*$, have some 
similarities, like the cut-off values and Gaussian-like, fast-decaying tails. A 
detailed 
assessment, however, requires further analysis, which we plan to publish 
elsewhere. 
\section{Summary}
\label{sec:sum}
We have studied an attachment scheme for growing directed networks, in which a 
new node attaches to an existing node $i$ with a probability 
$\Pi(k_{i})\propto 1/k_{i}$, where $k_i$ is the number of outgoing links at 
$i$. The motivation behind this idea is the following. If the links supply 
resources to the nodes in the network, and if the nodes can get the resources 
only through other nodes, then a new node will prefer to attach to existing 
nodes with fewer outgoing links to minimize its competitors for resources. We 
obtained the 
steady-state outdegree distribution, \nk, which decays like $k\mu^k/\Gamma(k)$ 
for large $k$, 
where $\mu$ is a constant. In order to further understand the effects of the 
low-degree preferential attachment, we also studied an equivalent growth model 
with a constant probability of attachment, which yields an exponential 
outdegree distribution. We also implemented this mechanism in the cascade 
model. The low-degree preferential attachment caused a difference 
between in- and outdegree distributions in the modified cascade model 
(identical in the original model), as well as a significant depression in the 
tail of the outdegree distribution and a broadening of the tail of the 
indegree distribution. These results indicate that the main effect of the 
low-degree preferential attachment mechanism is the fast decay in the tails 
of the outdegree distributions.

Although we propose this scheme for growing transportation networks, we do not 
claim that this is the only factor that determines the topology of such 
networks. In fact, we do not believe that growing transportation networks 
can be realistically modeled using a semi-static growth scheme, in which only 
the new 
nodes are allowed to make new connections while the rest of the network is 
frozen. Our model also lacks some essential features like the conservation of 
energy or mass. As a result, the networks can grow indefinitely. Nevertheless, 
the scheme we propose could be taken as a first approximation to modeling a 
growing directed transportation network, in which the probability of  
attachment is a decreasing function of the number of 
outgoing links.

\section*{Acknowledgments}
We appreciate helpful correspondence with S. Redner and P. L. Krapivsky. We 
also would like to thank an anonymous referee for providing important 
suggestions. This research was supported by U.S.\ National Science Foundation 
Grant Nos. DMR-0240078 and DMR-0444051, and by Florida State University 
through the School of Computational Science, the Center for Materials 
Research and Technology, and the National High Magnetic Field Laboratory.

\begin{figure}[t]
\centering \vspace{0.6truecm} 
\includegraphics[width=0.6\textwidth]{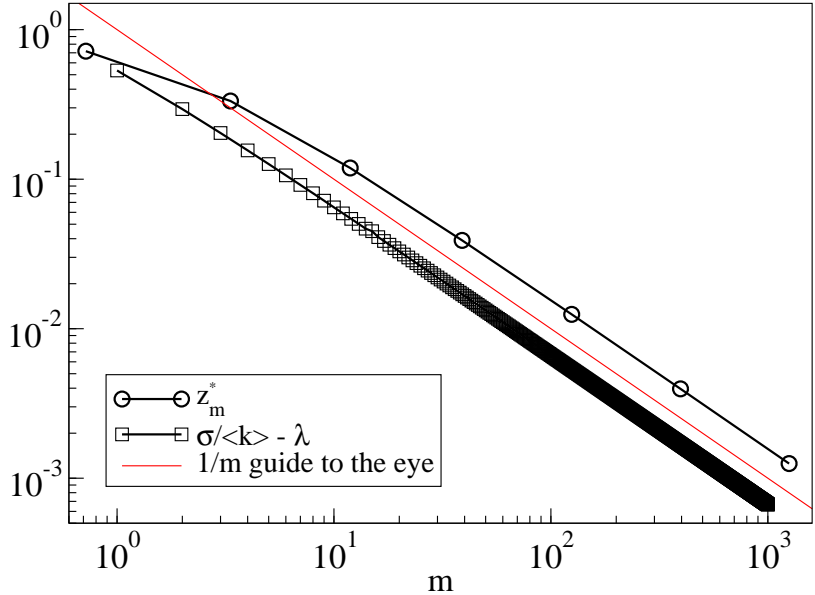}

\caption{(Color online) The $m$ dependence of $z_m^*$ and the convergence of 
the ratio of 
the standard deviation of $n_k^*$ to its mean, $\sigma/\langle k\rangle$, vs 
$m$. The graph shows that the standard deviation is proportional to the mean 
for large $m$. The ratio $\sigma/\langle k\rangle$ converges to 
$\lambda\approx 0.522$ like $1/m$. The $z_m^*$ vs $m$ plot confirms that 
$z_m^*\propto 1/m$ for large $m$.} 
\label{fig:converge}
\end{figure}
\begin{figure}[t]
\centering \vspace{0.6truecm} 
\includegraphics[width=0.6\textwidth]{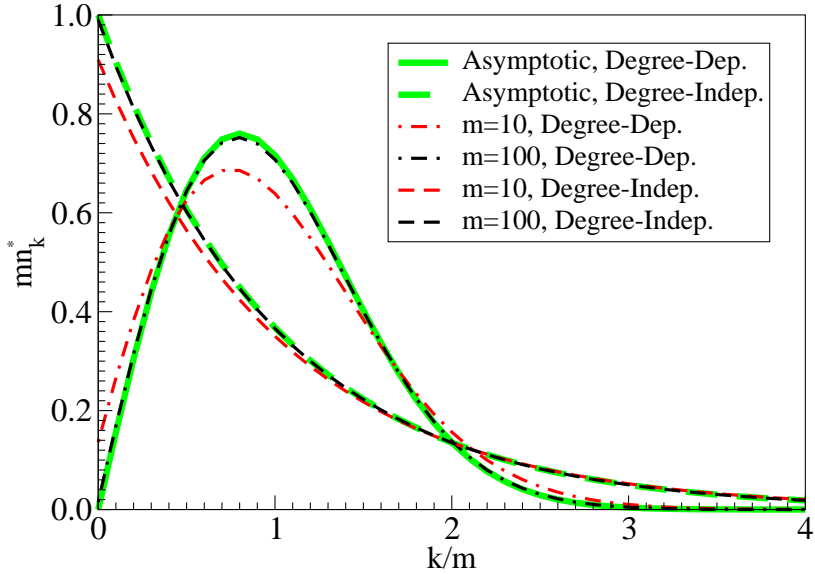}

\caption{(Color online) Scaled degree distributions (thin curves), $mn_k^*$ 
vs $k/m$, for the degree-dependent and 
degree-independent models (each for $m=$10 and 100), shown along with the 
asymptotic 
scaling functions (heavy curves), $p(u)=\frac{\pi 
u}{2}\exp(-\pi u^2/4)$ and $q(u)=\exp(-u)$ with $u=k/m$. For large values of 
$m$, the scaled distributions collapse onto their respective asymptotic 
scaling functions. Scaled distributions are obtained using analytical forms, 
Eq.~(\ref{eq:fullnk}) and Eq.~(\ref{eq:constnk}).} 
\label{fig:collapse}
\end{figure}

\appendix
\section{Numerical Calculation of $\zstar$}
\label{sec:zstar}
We use numerical methods to determine the values of $z_m^*$ for different $m$. 
Substituting Eq.~(\ref{eq:fullnk}) into Eq.~(\ref{eq:z}), yields a
self-consistent equation for $\zstar$:
\begin{equation}
\label{eq:zprod}
z_m^*= \sum_{k=0}^{\infty}
\frac{(\newzeta)^{k}\Gamma \big(1+\newzeta\big)}{\Gamma 
\big(k+2+\newzeta\big)} 
\;,
\end{equation}
which can also be written as
\begin{equation}
\label{eq:znumeric}
\zstar = \frac{\Phi(1,2+\newzeta;\newzeta)}{1+\newzeta}
\;,
\end{equation} 
where $\Phi(1,2+\newzeta;\newzeta)$ is the degenerate hypergeometric function 
defined as $\Phi(a,b;z)=\sum_{n=0}^{\infty}{\frac{(a)_n z^n}{(b)_n n!}}$, 
where $(a)_n=a(a+1)(a+2)\ldots(a+n-1)$, with $(a)_0=1$~\cite{GradshteynHyper}. 
The right-hand side of Eq.~(\ref{eq:znumeric}) is a smooth function of 
$\zstar$. We obtain $z_m^*$ by solving Eq.~(\ref{eq:znumeric}) numerically for 
$\newzeta$ \footnote{Here, one can also use the series representation in 
Eq.~(\ref{eq:zprod}) to obtain an approximate analytical solution. However, 
slow convergence of this series makes this method very tedious, and much less 
accurate.}. For $m=1$, 3, and 5, we obtained $z_m^*=0.6419$, 0.3582 and 0.2467, 
respectively, which we used to plot the outdegree distribution, 
Eq.~(\ref{eq:fullnk}), in Fig.~\ref{fig:nkplot}. The $m$-dependence of $z_m^*$ 
is shown in 
Fig.~\ref{fig:converge}. As seen in the figure, $z_m^*$ is proportional to 
$m^{-1}$ for large $m$. 


\section{Theoretical Calculation of the Scaling Function for the Growth Model 
with Low-degree Preferential Attachment}
\label{sec:app}
We scale $n_k^*$ by substituting $u=k/m$ and $p(u|m)=mn_{mu}$ in 
Eq.~(\ref{eq:nk}):
\begin{equation}
p(u|m)=\frac{(mu+1)}{z_m^*u[m/z_m^* + mu +1]}p(u-m^{-1}|m) \;.
\end{equation} 
By taking the logarithm of both sides, we obtain
\begin{equation}
\ln p(u|m)- \ln p(u-m^{-1}|m)=\ln\frac{1+1/mu}{1+z_m^*u+z_m^*/m}\;.
\end{equation} 
We convert the left-hand side to a derivative, and expand the right-hand side 
into a series assuming that $1/mu, z_m^*u$, and $z_m^*/m \rightarrow 0$ as 
$m\rightarrow \infty $:
\begin{equation}
\frac{1}{m} \frac{\partial}{\partial u}\ln p(u|m)=\frac{1}{mu}-
\frac{\zstar}{m}-\zstar u \;.
\end{equation}
This equation has a solution of the form
\begin{equation}
p(u|m)=Cu\exp (-z_m^*[u+mu^2/2]) \;.
\end{equation}
We require that $p(u|m)$ should be normalizable with $\langle u\rangle =1$ 
(recall that $\langle k\rangle=m$).
In the limit $m\rightarrow \infty $, this requires that $z_m^*=a/m$, where $a$ 
is a constant. Using these conditions, we normalize $p(u|\infty)\equiv 
p(u)=Cu\exp(-au^2/2)$, obtaining $C=\pi/2$ and $z_m^*=\pi/2m$. The 
asymptotic function, $p(u)$, is given in Eq.~(\ref{eq:scaling}) and shown 
in Fig.~\ref{fig:collapse}. The excellent agreement with $mn_k^*$ vs $k/m$ 
confirms our results.

This result also implies that the ratio of the standard deviation 
of $n_k^*$ to its mean, $\sigma/\langle k\rangle$, 
converges to a constant as $m$ increases. Therefore, $\sigma$ becomes 
proportional to $m$, since $\langle k\rangle$ is just $m$~\footnote{The proof 
of $\langle k\rangle=m$ is very simple: $\langle 
k\rangle=\sum{k_i}/N=\sum{k'_i}/N=Nm/N=m$.}. Fig.~\ref{fig:converge} shows 
that $\sigma/\langle k\rangle=\lambda +O(1/m)$ for large $m$. We obtain 
$\lambda=\lim_{m\rightarrow\infty }\sigma/\langle k\rangle$ using $p(u)$:
\begin{equation}
\lambda = \frac{\sqrt{\langle u^2\rangle-\langle u\rangle^2}}{\langle 
u\rangle}=(4/\pi-1)^{1/2}\approx 0.522\;.
\end{equation}


\section{Theoretical Calculation of Outdegree Distribution for the Growth Model 
with Degree-Independent Probability of Attachment}
\label{sec:const}
We follow the same method we used to derive Eq.~(\ref{eq:fullnk}). Taking the 
probability of attachment independent of in- and outdegree, and also identical 
for all nodes, $\Pi=1/N_{\rm T}$, we obtain the rate equations for the outgoing 
links:

\begin{equation}
\frac{dN_k}{dt} = m\frac{N_{k-1}}{N_{\rm T}} - m\frac{N_k}{N_{\rm T}} 
\qquad\textrm{for }k>0
\;.
\end{equation}
Substituting  $n_{k}N_{\rm T}$ for $N_{k}$, and using $dN_{\rm T}/dt=1$, yields 
equations for the density of nodes:
\begin{equation}
N_{\rm T}\frac{dn_k}{dt} = mn_{k-1} - (m+1)n_k \;.
\end{equation} 
In the steady state, we have
\begin{equation}
(m+1)n_k^* = mn_{k-1}^* \;.
\end{equation} 
The solution of this equation yields the steady-state outdegree distribution:
\begin{equation}
n_k^* = n_0^*c^k \;, 
\label{eq:constnk}
\end{equation} 
where $c=m/(1+m)$. Imposing the normalization condition, 
$\sum_k^\infty{n_k^*}=1$, we find, $n_0^*=1-c=1/(1+m)$. 

We can also derive the scaling form of this distribution for comparison with 
the scaling form of Eq.~(\ref{eq:fullnk}) in Appendix~\ref{sec:app}. 
Multiplying both sides of Eq.~(\ref{eq:constnk}) by $m$, we get
\begin{equation}
mn_k^* = \left(\frac{m}{1+m}\right)^{k+1}=[(1+1/m)^m]^{-\frac{k+1}{m}} 
\;.
\end{equation}
In the limit $m\rightarrow \infty$, this equation yields the scaling function
\begin{equation}
q(u)=mn_k^* = \exp(-u) \;,
\end{equation}
with $u=k/m$. $q(u)$ is shown in Fig.~\ref{fig:collapse}.

As a result, the asymptotic ratio  
$\sigma/\langle k\rangle$ for the outdegree distribution is equal to unity, 
a characteristic of the 
exponential distribution, compared to $ \lambda\approx 0.522$ in 
degree-dependent case.

%

\end{document}